\documentclass[apj]{emulateapj}
\usepackage{apjfonts}

\shorttitle{Blue Stragglers in the lowest stellar density systems}
\shortauthors{Santana et al.}

\begin{document}

\title{A Megacam Survey of Outer Halo Satellites. II. Blue Stragglers in the Lowest
Stellar Density Systems \altaffilmark{1}}
\author{
Felipe\ A.\ Santana\altaffilmark{2,3},
Ricardo\ R.\ Mu\~noz\altaffilmark{2,3},
Marla\ Geha\altaffilmark{3},
Patrick\ C\^ot\'e\altaffilmark{4},
Peter\ Stetson\altaffilmark{4},
Joshua\ D.\ Simon\altaffilmark{5}, \&
S.\ G.\ Djorgovski\altaffilmark{6}
}

\altaffiltext{1}{Based on observations obtained at the Canada-France-Hawaii Telescope 
(CFHT) 
which is operated by the National Research Council of Canada, the Institut National des 
Sciences de l'Univers of the Centre National de la Recherche Scientifique of France, 
and the University of Hawaii.}

\altaffiltext{2}{Departamento de Astronom\'ia, Universidad de Chile, Camino El 
Observatorio 1515, Las Condes, Santiago, Chile (fsantana@das.uchile.cl, 
rmunoz@das.uchile.cl)}

\altaffiltext{3}{Astronomy Department, Yale University, New Haven, CT 06520, USA}

\altaffiltext{4}{Herzberg Institute of Astrophysics, National Research Council of Canada, 
Victoria, BC, V9E 2E7, Canada}

\altaffiltext{5}{Observatories of the Carnegie Institution of Washington, 813 Santa 
Barbara St., Pasadena, CA 91101, USA}

\altaffiltext{6}{Astronomy Department, California Institute of Technology, Pasadena, CA, 
91125, USA}

\begin{abstract}

We present a homogeneous study of blue straggler stars across ten outer halo 
globular clusters, three classical dwarf spheroidal and nine ultra-faint galaxies based on 
deep and wide-field photometric data taken with MegaCam on the Canada-France-Hawaii
Telescope. We find blue straggler stars to be ubiquitous among these Milky Way satellites. 
Based on these data, we can test the importance of primordial binaries or multiple systems
on blue straggler star formation in low density environments.
For the outer halo globular clusters we find an anti-correlation between the specific 
frequency of blue straggler and absolute magnitude, similar to that previously 
observed for inner halo clusters.
When plotted against density and encounter rate, the frequency of blue stragglers are well 
fitted by single trends with smooth transitions between dwarf galaxies and globular 
clusters, which points to a common origin for their blue stragglers.
The fraction of blue stragglers stays constant and high in the low encounter rate regime 
spanned by our dwarf galaxies, and decreases with density and encounter rate in the range 
spanned by our globular clusters.
We find that young stars can mimic blue stragglers in dwarf galaxies
only if their ages are $2.5\pm0.5$\,Gyr and they represent
$\sim1$--$7$\% of the total number of stars, which we deem highly unlikely.
These results point to mass-transfer or mergers of primordial binaries or multiple systems 
as the dominant blue straggler formation mechanism in low density systems.

\end{abstract}


\section{Introduction}

Blue stragglers are stars coeval with a given stellar population, but 
positioned 
blueward and above its main sequence (MS) turnoff, thus mimicking a younger population.
They were first observed by \citet{sandage53a} in the globular cluster M3 as an apparent
extension of the classical MS (see for example \citealt{bailyn95a} for a review on 
blue stragglers). 
Since globular clusters have traditionally been considered single stellar populations, 
stars located blueward and above the MS turnoff in its color-magnitude diagram (CMD)
 should have evolved out of the MS into a post hydrogen-burning phase.
 In this context, the  existence of blue straggler stars challenges our current understanding of 
stellar evolution.
To inhabit a hotter and more luminous region in the CMD,
these stars must have increased their original masses and, in the process,
renewed their fuel for nuclear reactions.

Since the discovery made by Sandage, blue stragglers have been found in practically all 
Galactic globular clusters, and several formation mechanisms have been proposed. 
Early on, blue stragglers as single stars were proposed, either massive young stars due 
to recent star formation, or stars in a post-helium flash evolutionary phase where 
hydrogen rich material has sunk to the core \citep{rood70a,conti74a}, but these were later
discarded \citep[e.g.,] []{nemec87a,nemec89a}.
At present, the leading blue straggler star formation mechanisms are stellar mergers 
produced by direct stellar collisions 
\citep[hereafter \textit{collisional blue stragglers}, e.g.,][]{hills76a,leonard89a} and 
mass-transfer or mergers in primordial
binary or higher order systems 
\citep[hereafter \textit{binary blue stragglers}, e.g.,][]{mccrea64a,knigge09a, perets09a}. 
Several studies have shown that some blue stragglers are indeed binary systems, by 
measuring photometric variability in these stars
\citep{jorgensen84a,mateo90a,mateo95a,nemec95a}.
On the other hand, triples have been claimed to be particularly
relevant in blue straggler formation in low density environments \citep{leigh11b}.
Triples could form blue stragglers through mechanisms like Kozai cycles and 
tidal friction
\citep{perets09a} or triple evolution dynamical instabilities \citep{perets12a}.
The importance of collisions involving triple stars in forming blue stragglers was confirmed
by \citet{geller13a} through $N-$body modeling of the old open cluster NGC~188.
\citet{bailyn95a} argued that both mechanisms (binary and direct collisions)
are likely to be at work in globular clusters, a view shared by several studies \citep[e.g.,][]
{hurley01a,mapelli06a,dalessandro08a,ferraro09a} with their relative importance being a 
function of cluster mass \citep{davies04a}, dynamical evolution and physical conditions
 of the environment \citep{piotto04a,knigge09a,leigh11a}.

To investigate the relative importance of the two formation mechanisms, 
several correlations between the observed fraction of blue straggler stars and host 
properties have been explored. 
For example, the fraction of blue stragglers can be plotted as a function of density or 
encounter rate. 
If the fraction of blue stragglers grows with density, then collisions
might be the dominant formation mechanism. If instead we find less blue
stragglers in denser systems, collisions between stars might 
prevent blue straggler formation, either by separating primordial binaries or 
disrupting multiple star systems.
Perhaps the most notable result in this context is the one
reported by \citet{piotto04a}, who found that blue straggler
specific fraction declines with increasing luminosity or mass.
The interpretation of this anti-correlation is that the current fraction
 of binary stars, from which blue stragglers would form, would be lower for larger
and denser systems \citep{leigh11a,sollima08a,davies04a}. Although, in the magnitude
 range spanned by Piotto's globular cluster sample, $-10<M_{\rm{V}}<-6$, a significant 
contribution from collisionally formed blue straggler stars cannot be discarded.

In addition to globular clusters, blue stragglers have been detected in a variety
of low density environments
such as Galactic dwarf spheroidal (dSph) galaxies 
\citep[e.g.,][]{momany07a}, loose stellar clusters \citep[e.g.,][]{geller11a, sollima08a}, 
the Milky Way's bulge \citep{clarkson11a} and even the Galactic field 
\citep[e.g.,][]{stetson91a,glaspey94a,preston00a, carney01a, carney05a}.
Currently, the number of studies in dSph galaxies is rather limited, and in most cases
 they cover one or two galaxies (see \citealt{mapelli07a} for Draco and Ursa Minor, 
 \citealt{hurleykeller99a} and \citealt{monkiewicz99a} for Sculptor, and \citealt{mateo95a} 
for Sextans). Only one study to date presents a systematic study among classical dSph 
galaxies \citep{momany07a}.

\begin{deluxetable*}{llrrrrrlrr}
\tablecolumns{8}
\tablewidth{17.5cm}
\tabletypesize{\scriptsize}
\tablecaption{Blue Straggler Stars and RGB Counts for All Satellites}
\tablehead{ \colhead{Object} & \colhead{Type} & 
\colhead{BSS$_{\rm{T}}$\tablenotemark{1}} & 
\colhead{Norm BSS$_{\rm{C}}$\tablenotemark{2}} &
\colhead{RGB$_{\rm{T}}$\tablenotemark{3}} & 
\colhead{Norm RGB$_{\rm{C}}$\tablenotemark{4}} & 
\colhead{F$_{\rm{RGB}}^{\rm{BSS}}$\tablenotemark{5}} &
\colhead{n$_{\rm{rh}}$[stars pc$^{\rm{-3}}$]\tablenotemark{6}} &
\colhead{$\frac{r_{\rm{cont,inn}}}{r_{\rm{half}}}$\tablenotemark{7}} &
\colhead{$\frac{r_{\rm{cont,out}}}{r_{\rm{half}}}$\tablenotemark{8}}}
\startdata
Bo\"otes I& UFDG&71&16.19&230 & 77.73 & $0.36\pm0.07$ & 2.1$\times 10^{-3}$& 5.0&6.0\\
Bo\"otes II& UFDG&4&1.20&18 & 5.00 & $0.26\pm0.17$ & 5.9$\times 10^{-3}$& 4.0&6.0\\
Coma Berenices&UFDG&10&4.00&21 & 8.83 & $0.49\pm0.33$ & 4.6$\times 10^{-3}$& 5.0&7.0\\
Canes Venatici I& UFDG&343&91.98&842&40.37& $0.31\pm0.03$ &6.0$\times 10^{-4}$& 4.8&5.8\\
Hercules&UFDG&35&9.43&127&27.57& $0.26\pm0.07$ &4.1$\times 10^{-4}$&6.0&8.0\\
Segue 2& UFDG&3&0.60&13 & 3.57 & $0.26\pm0.22$ & 7.4$\times 10^{-3}$&5.0&6.2\\
Ursa Major I& UFDG&19&6.36&56& 11.59 & $0.28\pm0.12$ & 3.8$\times 10^{-4}$& 5.5&7.0\\
Ursa Major II& UFDG&23&3.85&79& 16.42 & $0.31\pm0.09$ & 1.2$\times 10^{-3}$& 5.0&7.0\\
Willman 1& UFDG&3&0.21& 4 & 0.72 & $0.85\pm0.75$ & 2.2$\times 10^{-2}$& 5.0&8.0\\
Draco& dSph&422&17.82&1600 & 76.73 & $0.27\pm0.02$ & 1.2$\times 10^{-2}$& 5.0&6.0\\
Sextans& dSph&521&65.49&1822& 260.35& $0.29\pm0.02$ & 1.3$\times 10^{-3}$&4.0&5.0\\
Ursa Minor& dSph&514&14.00&1743 & 47.36 & $0.30\pm0.02$ & 3.3$\times 10^{-3}$&5.0&6.5\\
NGC~5694&GC& 11 & 0.15 & 247 & 6.97 & $0.05\pm0.01$  & 7.5$\times 10^{0}$& 15.0&19.0\\
NGC~6229&GC& 11 & 0.19 & 318 & 1.66 & $0.03\pm0.01$ &1.5$\times 10^{1}$& 14.0&18.0\\
NGC~7006&GC& 19 & 0.12 & 263 & 9.06 & $0.07\pm0.02$  & 6.1$\times 10^{0}$&14.0&18.0\\
NGC~7492&GC& 8 & 0.08 & 83 & 0.75 & $0.10\pm0.04$ & 1.4$\times 10^{1}$&16.0&20.0\\
Eridanus &GC&12 & 0.25 & 45 & 0.39 & $0.26\pm0.09$ &2.7$\times 10^{-1}$& 20.0&28.0\\
Palomar~3&GC& 11 & 0.15 & 65 & 0.44 & $0.17\pm0.06$ & 8.5$\times 10^{-1}$& 20.0&30.0\\
Palomar~4&GC& 10 & 0.12 & 102 & 0.44 & $0.10\pm0.03$ & 1.5$\times 10^{0}$& 20.0&30.0\\
Palomar~13&GC& 8 & 0.12 & 20 & 0.75 & $0.41\pm0.18$ & 1.4$\times 10^{0}$& 16.0&22.0\\
Palomar~14&GC& 15 & 0.75 & 91 & 5.14 & $0.17\pm0.05$ & 1.6$\times 10^{-1}$& 12.0&16.0\\
Palomar~15&GC& 20&1.32&150&18.25& $0.14\pm0.04$ &6.5$\times 10^{-1}$&12.0&16.0\\[1ex]
\enddata
\tablenotetext{1}{Blue straggler stars measured in the system region}
\tablenotetext{2}{Blue straggler stars measured in the contamination region normalized by area}
\tablenotetext{3}{RGB stars measured in the system region}
\tablenotetext{4}{RGB stars measured in the contamination region normalized by area}
\tablenotetext{5}{Specific fraction of blue stragglers as measured from equation (1)}
\tablenotetext{6}{Stellar density measured within the half-light radius of the system}
\tablenotetext{7}{Inner radius of contamination region normalized to the half-light radius}
\tablenotetext{8}{Outer radius of contamination region normalized to the half-light radius}
\label{blue straggler star frequencies}
\end{deluxetable*}

On the other hand, information regarding blue straggler stars in ultra-faint dwarf (UFD) galaxies
is extremely scarce.
The UFDs correspond to a recent population of dark matter-dominated satellites found
in the Milky Way halo as stellar overdensities in the Sloan Digital Sky Survey
photometric catalog
\citep[SDSS; e.g.,][]{willman05a, belokurov06a, belokurov07a, belokurov10a, zucker06a, 
irwin07a}, with total luminosities even lower than those of the previously known dSphs,
 ranging from $M_{\rm{V}}\sim-8$ down to $M_{\rm{V}}=-1.5$ for the faintest system yet 
found (Segue~1). 
These extreme low luminosity, low surface brightness systems represent a new opportunity
 to study blue straggler stars in extremely low stellar density environments.

In this article, we use a new survey carried out with the Canada-France-Hawaii
 Telescope (CFHT, R. R. Mu\~noz, in preparation). This survey is aimed at obtaining
wide-field photometry of all bound stellar over-densities in the outer halo
(i.e., with Galactocentric distances greater than $25$\,kpc.).
Here we present a homogeneous analysis of the blue straggler star populations
 of most of these Milky Way  halo satellites, to study the characteristics of blue straggler 
stars in the lowest stellar density systems. The analysis includes globular
clusters, dSphs, as well as the first systematic study of blue stragglers in the UFDs.

Collisions involving single, binary or triple stars in our systems show typical times of
 occurrence that are large compared to the blue straggler lifetime. Therefore, except for a
few of our highest density systems, the blue stragglers in our Galactic satellites might not
be explained by collisions of any type.
Moreover, the densest objects in our sample, the outer halo clusters,
are as a group, fainter and on average ten times bigger than their inner halo counterparts. 
Our goal is to investigate the presence/absence of blue straggler stars in the most diffuse 
stellar systems and study the influence of the environment on their blue straggler star 
populations.
The  article is organized as follows:
In Section~2 we describe the photometric catalog and the sample of satellites used in
 this study. 
We also detail how blue straggler stars are selected and their counts normalized to 
Red-Giant-Branch (RGB) stars. 
In Section~3 we present our results, including modeling of young populations that could be
mimicking blue straggler stars in dwarf galaxies.
In Section~4 we discuss our results for each type of satellite. Finally, a brief summary is 
presented in Section~5.

\section{Data and Blue Straggler Selection}

We analyzed the blue straggler star population of $22$
outer halo (beyond $R_{\rm{G}}=25$\,kpc) satellites: ten globular
clusters, three of the classical dSph galaxies (Draco, Ursa Minor and Sextans) and
nine UFDs. 
Data for all objects were obtained using the CFHT MegaCam imager, and it represents
a subsample of a larger 
program aimed at obtaining wide-field photometry of all bound stellar over-densities
in the outer halo (R.~R.~ Mu\~noz et al., in preparation). First results from this survey were
 presented in \citet{munoz12b}, who reported the discovery of a very low luminosity star
 cluster at a distance of $\sim45$\,kpc along the line of sight to the Ursa Minor dSph galaxy.
 From the entire sample of objects in this survey, the
$22$ systems considered here correspond to the ones where blue straggler stars could be reliably
selected in the color-magnitude diagram. In the excluded systems, extreme low star counts,
severe overcrowding or photometry issues prevented  a robust blue straggler discrimination,
and would have added large systematic errors to the analysis.

For each system, six dithered exposures in the CFHT $g-$ and $r-$band were
 taken in mostly dark conditions.
A standard dithering pattern was chosen from the MegaCam operation options 
to cover all gaps between chips.
The exposure times varied between objects, but in all cases
reached at least one magnitude below the MSTO of the 
satellite. Details of the observing log (i.e., exposure time, seeing conditions, etc)
will be given elsewhere.

\begin{figure*}
\plotone{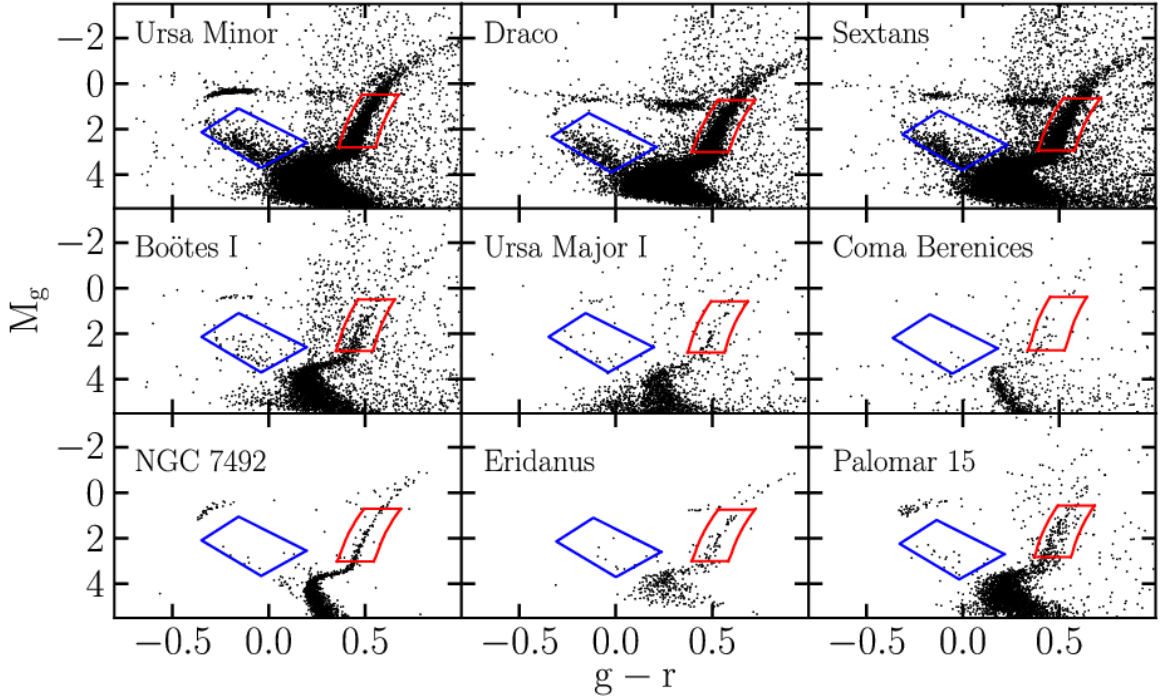}
\caption{$(g-r)$ vs $M_{\rm{g}}$ extinction corrected color-magnitude diagrams of stars in 
9 different satellites from our sample, obtained at the Canada-France-Hawaii Telescope. 
Classical dwarf galaxies are shown in the top panels, 
ultra-faint dwarf (UFD) galaxies are shown in the middle panels and three of the globular 
clusters
in our sample are shown in the bottom panels. Boxes where blue straggler and red giant
 branch stars were counted are shown for each case.}
\label{diff_CMDs}
\end{figure*}

Data from MegaCam are pre-processed prior to release using the ``ELIXIR'' package
 \citep{magnier04a}, which includes bias subtraction, flat fielding, bad pixel correction 
and preliminary solutions for photometry and astrometry. We carried out subsequent data 
analysis
using the DAOPHOT/Allstar/ALLFRAME package \citep{stetson94a}, following the
method detailed in \citet{munoz10a}. To remove non-stellar objects and spurious 
detections, we used the DAOPHOT {\it Chi} and $sharp$ parameters.
We selected sources with a $Chi < 5$ and $-0.4<sharp<0.4$.
In addition, we only used detections with photometric uncertainties smaller than
 $0.1$ magnitudes in both bands. Finally, we calibrated our observations using SDSS
data, which allowed us to obtain our final calibrated magnitudes translated into Sloan
$g-$ and $r-$ bands.
We note that overcrowding affects only a few objects since most
outer halo clusters are, on average, more extended and less luminous than their
inner halo counterparts, and therefore have lower stellar densities.
In the cases where overcrowding could have been a problem, the regions where completeness
 was below $50$\%, for magnitudes brighter than that of the MSTO, were always
more than a factor of $50$ smaller than the total region studied. Therefore, even
for these systems overcrowding produces a negligible effect on our results.
In Figure~\ref{diff_CMDs}  we show the CMDs of all the sources selected as stars for 
different globular clusters, classical dwarf galaxies and UFD galaxies in our sample.

\begin{figure*}
\plotone{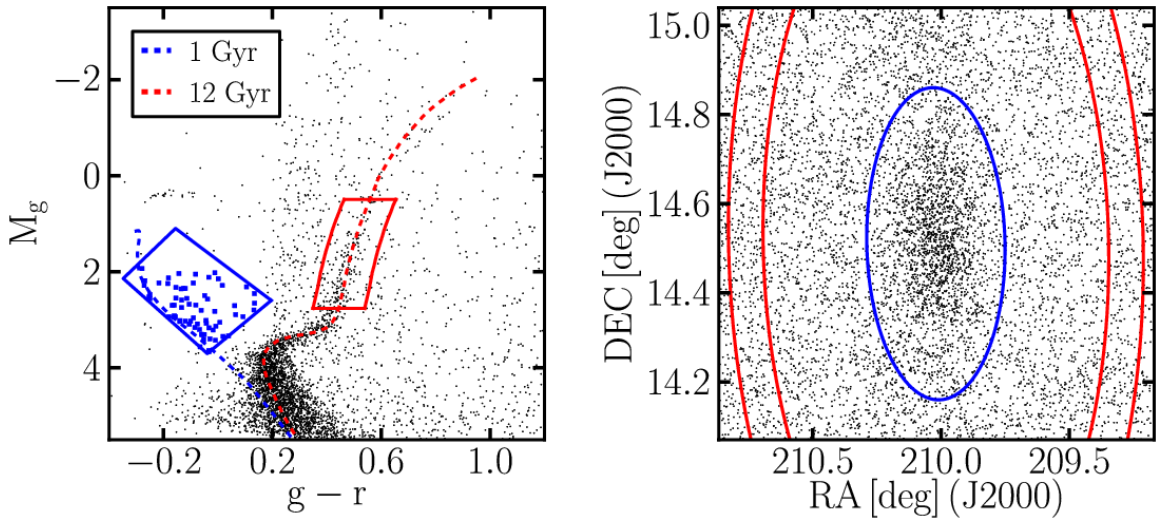}
\caption{Stars in the Bo\"otes I field. 
{\it Left}: Bo\"otes I $(g-r)$ vs $M_{\rm{g}}$, extinction corrected color-magnitude diagram. 
Blue and red lines show, respectively, Padova isochrones of $1$ and $12$\,Gyr, with a 
metallicity of [Fe/H]=$-2.1$. The blue box shows the color-magnitude diagram region were
 blue straggler stars were counted, while the RGB region is delimited by the red box. 
Stars considered as blue stragglers are shown as blue squares.
{\it Right}: Star map of the Bo\"otes I UFD.  The blue curve shows the limit for the
 system region at twice the half-light radius, while the red annuli show the limits for the
 contamination region, at $5$ and $6$ times the half-light radius respectively.}
\label{BooI_CMD}
\end{figure*}

To select blue straggler stars, we defined a box in the dereddened $(g-r)$ vs 
$M_{\rm{g}}$ diagram of each object.
The size and shape of this box was chosen to maximize the blue straggler counts while
at the same time minimizing contaminants from other stellar populations. We determined
the distance from the blue straggler box to the MSTO position based on typical main-sequence
widths and photometric uncertainties. Likewise, the bright side of the blue straggler box was
chosen based on typical blue horizontal branch extensions. 
In Figure~\ref{BooI_CMD} we show our blue straggler star selection criteria for the ultra 
faint dwarf galaxy Bo\"otes I as an illustration. 
The coordinates of the four points in the CMD that define the blue straggler star\footnote{It
 is worth noting that we denominate  {\it blue stragglers} what in principle can 
also be young stars. However, in Section~3 we show that the vast majority of our blue straggler 
``candidates'' are unlikely to be young stars. We therefore, use the term blue straggler for
 every star that falls inside the box in the CMD described above.} 
 box for each object, relative to the MSTO position had typical $[\Delta(g-r),\Delta g]$ 
values of ($-0.245$,$-0.405$), ($-0.554$,$-1.969$), ($-0.362$,$-3.012$) and ($-0.010$,
$-1.513$).
Small shifts, with typical values of $0.015$ in color and $0.09$ in 
magnitude, were applied in some objects to move the entire box. These shifts are mainly
caused by uncertainties in the position of the MSTO and the intention of avoiding regions
significantly contaminated in certain objects.
Varying the bright and dim side of our blue straggler box, along with applying the 
shifts, does not significantly change our results since these variations were proven to be 
small compared to our random errors.

The data were dereddened for all satellites using values for $E(B-V)$ taken from
\citet{schlegel98a}.
These values
were translated into Sloan filter extinctions A$_g$ and A$_r$, using the transformations 
of \citet{schlafly11a}.
Absolute magnitudes were derived using distance values from the literature\footnote{ 
Distances to all our clusters were taken from \citet{harris10a}, for dwarf galaxies the 
following references were used: \citet {dallora06a} for Bo\"otes I; \citet {walsh08b} for
Bo\"otes II; \citet{musella09a} for Coma Berenices; \citet{kuehn08a} for Canes Venatici I;
\citet{bonanos04a} for Draco; \citet{musella12a} for Hercules; \citet{belokurov09a} for
Segue 2; \citet{lee03a} for Sextans; \citet{garofalo13a} for Ursa Major I; \citet{dallora12a}
for Ursa Major II; \citet{carrera02a} for Ursa Minor and \citet{willman05b} for Willman 1.}.
We derived metallicities by fitting Padova isochrones to the main old population. 

\begin{figure}
\plotone{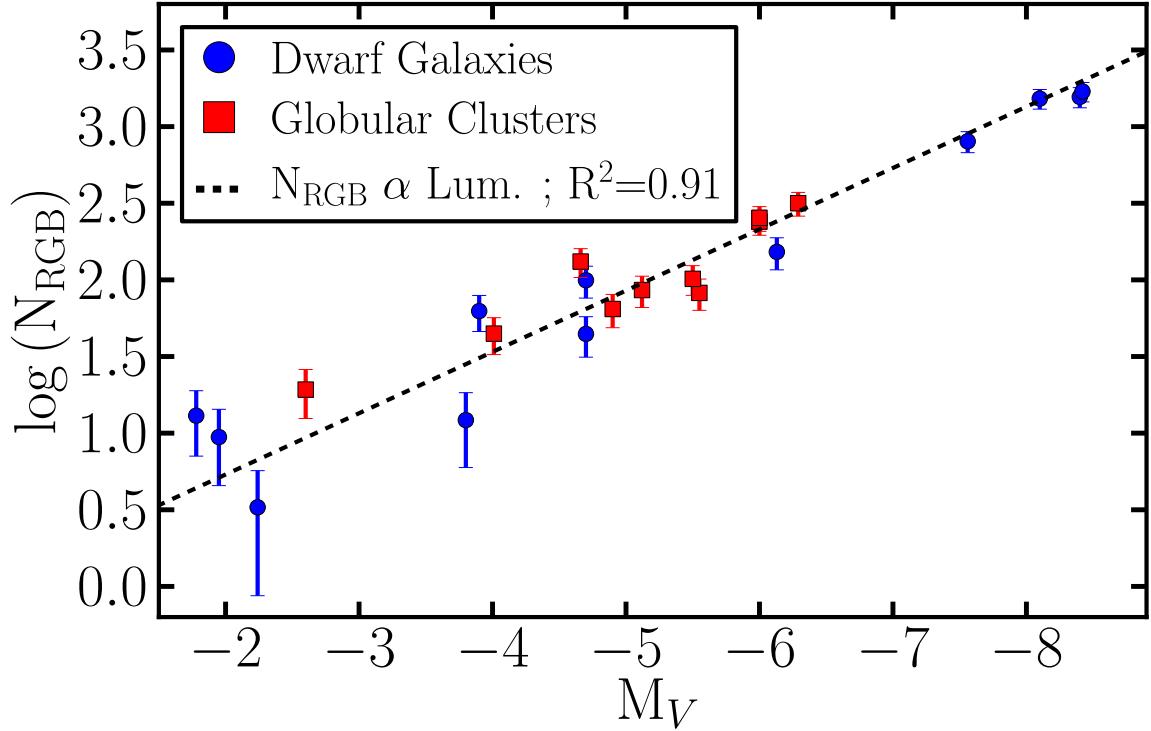}
\caption{Red giant branch star numbers are plotted against the absolute magnitude of each
system. The figure shows that our normalization population numbers, are
 directly proportional to the total luminosity of the system. Dashed black line shows a linear 
relation between luminosity and red giant branch star counts, which has a correlation factor 
of $0.91$ with our data.
}
\label{rgb_linearity}
\end{figure}

To compare the number of blue straggler stars among systems with different absolute 
magnitude, a common practice \citep[e.g.,][]{piotto04a, leigh07a} is to normalize the blue
straggler star counts to those of another sub-population, typically RGB or blue horizontal
branch stars.
For this study, we chose RGB stars since they are more numerous than blue horizontal
branch stars. This choice reduces shot noise due to low number of stars, a problem 
especially critical for the UFDs \citep{martin08b, munoz12a}. 
To avoid introducing significant bias when using RGB stars as a normalization population,
 we checked that the number of these stars grows linearly with luminosity.
 Figure~\ref{rgb_linearity} shows that RGB star counts are indeed proportional to the flux of
 the systems, with a correlation factor of $r^2=0.91$, confirming that RGB stars are good
tracers of total stellar luminosity or mass.
To select RGB stars, we defined a box centered on a $12$\,Gyr old Padova isochrone 
(with the appropriate metallicity for each system) $0.19$\,mag wide and located
between $2.4$ and $4.9$\,mag below the RGB tip. 
To count both blue straggler and RGB stars we used an elliptical region within 
$2$ times the half-light radius ($r_{\rm{h}}$) of the system. 
For each case, this region was defined by the ellipticity and position angle derived
in R.~R.~Mu\~noz et al. (in preparation), based on the same CFHT data.
To account for background/foreground contaminants, we counted sources in both the blue 
straggler and
RGB boxes, but in annuli at distances greater than $4\times r_{\rm{h}}$
from the center of the object.
An example is shown in the right panel of Figure~\ref{BooI_CMD} and the values of the
inner and outer annuli of the contamination region, in terms of $r_{\rm{h}}$, are shown in
Table~\ref{blue straggler star frequencies} for each object.

Once blue stragglers, RGB stars and contamination objects were selected, we defined the
specific fraction of blue stragglers as:
\begin{equation}
F_{\rm{RGB}}^{\rm{BSS}}=\frac{BSS_{\rm{s}}-BSS_{\rm{c}}}{RGB_{\rm{s}}-RGB_{\rm{c}}}
\end{equation}

\noindent where $BSS$ and $RGB$ mean `blue straggler star' and `red giant branch star' and 
$s$ and $c$ mean, respectively, `system' and `contamination'.

\section{Results}

\subsection{Blue Straggler Specific Fractions}

In Table~\ref{blue straggler star frequencies} we list the resulting blue straggler and RGB star 
counts, along with the corresponding blue straggler specific fractions, limits for the 
contamination region and the density of our clusters and galaxies.

\begin{figure}
\plotone{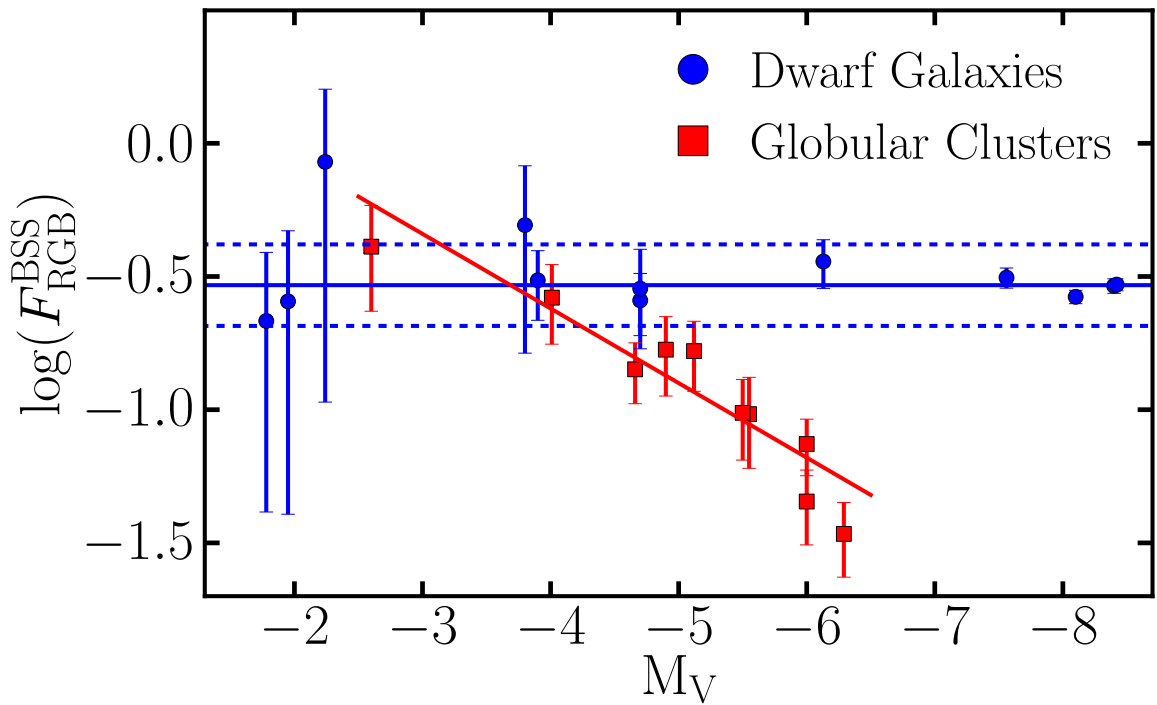}
\caption{Specific fraction of blue stragglers $F_{\rm{RGB}}^{\rm{BSS}}$ 
is plotted against absolute magnitude. 
A clear anti-correlation can be seen for clusters, while dwarf galaxies show a high and flat
 distribution.
The logarithm of the weighted mean of ${F_{\rm{RGB}}^{\rm{BSS}}}$ is shown as a solid blue line while dashed blue lines show the standard deviation around this value. 
Red solid line shows the fit for clusters corresponding to 
$\log({F_{\rm{RGB}}^{\rm{BSS}}})$ $\propto$ ($0.28\pm0.04$) $M_{V}$
}
\label{freq_magnitude}
\end{figure}

A surprising first result is that blue stragglers seem to be ubiquitous among dwarf galaxies, 
being present even in the most diffuse and least luminous systems. 
In Figure~\ref{freq_magnitude}, we plot $F_{\rm{RGB}}^{\rm{BSS}}$ against 
absolute magnitude, $M_{\rm{V}}$. 
This figure shows that the blue straggler star fraction distribution for galaxies is 
statistically consistent with being flat over a six magnitude range, 
with a weighted mean value of:

\begin{equation}
\label{average_freq_dwarfs}
\left. \overline{F_{\rm{RGB}}^{\rm{BSS}}}\right| \rm{_{dwarfs}} = 0.29\pm0.01
\end{equation}
 and a standard deviation of $0.17$.
In contrast, for globular clusters we see a well-defined anti-correlation between
$\log({F_{\rm{RGB}}^{\rm{BSS}}})$ and the absolute magnitude of the objects. 
The linear function fitted has the form:  
\begin{equation}
\label{eq:freq_mag_clusters}
\left. \log({F_{\rm{RGB}}^{\rm{BSS}}})\right|{\rm_{clusters}} = (0.28\pm0.04) M_V + 
(0.50\pm0.22).
\end{equation}
The uncertainties in the fitting parameters of this and all the forthcoming equations 
were estimated using Monte Carlo simulations. Each time we ran a simulation, we shifted 
the data by values consistent with the uncertainties in the measured frequencies
and then calculated the set of fitting 
parameters that corresponded to that shifted data sample. Then, for each fitting parameter, 
the uncertainty was determined as the standard deviation of the values obtained in the 
different runs.

This result is consistent with a similar anti-correlation found by \citet{piotto04a} for a
group of $56$ globular clusters, most of them in the inner halo, but in our study we have 
expanded the anti-correlation to clusters that are three magnitudes fainter. Even though
we used RGB stars as a normalization population and HB stars were used in 
\citet{piotto04a}, the slopes of the anti-correlations found in both studies are consistent
 within the errors.

\begin{figure}
\plotone{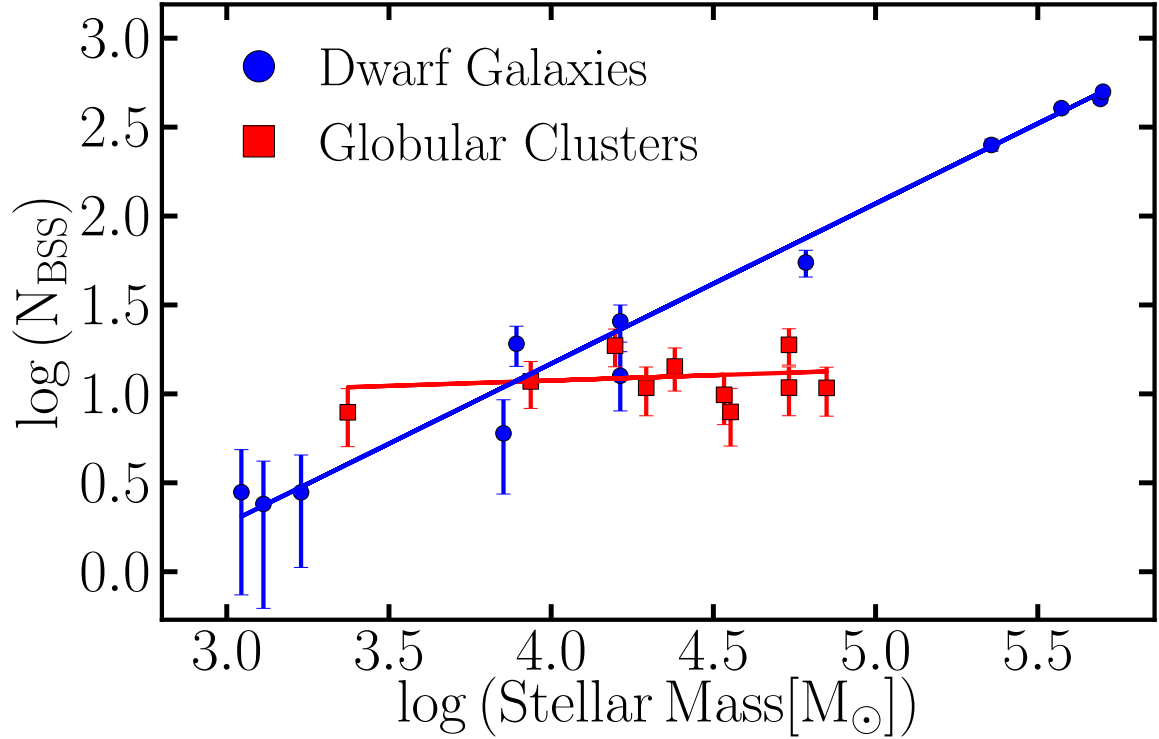}
\caption{The number of blue stragglers is plotted against the total stellar mass of each
 system. Fitting function for clusters is $\rm{\log{(N_{BSS})}}$
$\propto$ ($0.06\pm0.07$) $\rm{\log{(M)}}$ and is shown as a red line. Fitting function
for galaxies is $\rm{\log{(N_{BSS})}}$
$\propto$ ($0.90\pm0.04$) $\rm{\log{(M)}}$ and is shown as a blue line.
}
\label{number_mass}
\end{figure}

A different normalization method (first outlined in \citealt{knigge09a}) was also used to 
illustrate the dependence of blue straggler population sizes on the total population 
size of their hosts.
As seen in Figure~\ref{number_mass}, we correlated the number of blue stragglers 
observed with the total {\it stellar} mass of our systems. 
The linear fitting functions we obtained using this normalization were:
\begin{equation}
\label{number_mass_clusters}
\log{(N_{\rm{BSS}})} = (0.06\pm0.07) \log{(\rm{Mass})} + (0.8\pm0.3)
\end{equation}
\begin{equation}
\label{number_mass_dwarfs}
\log{(N_{\rm{BSS}})} = (0.90\pm0.04) \log{(\rm{Mass})} + (-2.4\pm0.2)
\end{equation}
for globular clusters and dwarf galaxies respectively.

Both correlations found here are equivalent to the results found before using the specific
frequency of blue stragglers. Blue straggler numbers in clusters increasing slowly with mass
is equivalent to an anti-correlation of $F_{\rm{RGB}}^{\rm{BSS}}$ and 
$M_{\rm{V}}$ like the one in equation~\ref{eq:freq_mag_clusters}. On the other hand, blue straggler
numbers in dwarf galaxies growing almost linearly with mass is equivalent to a nearly
constant distribution of $F_{\rm{RGB}}^{\rm{BSS}}$. Within the errors, 
equations~\ref{average_freq_dwarfs} and \ref{number_mass_dwarfs} point
to specific blue straggler fractions in dwarfs that are either independent of the absolute
 magnitude or that follow a shallow anti-correlation with absolute magnitude like the one
found by \citet{momany07a}.

\begin{figure}
\plotone{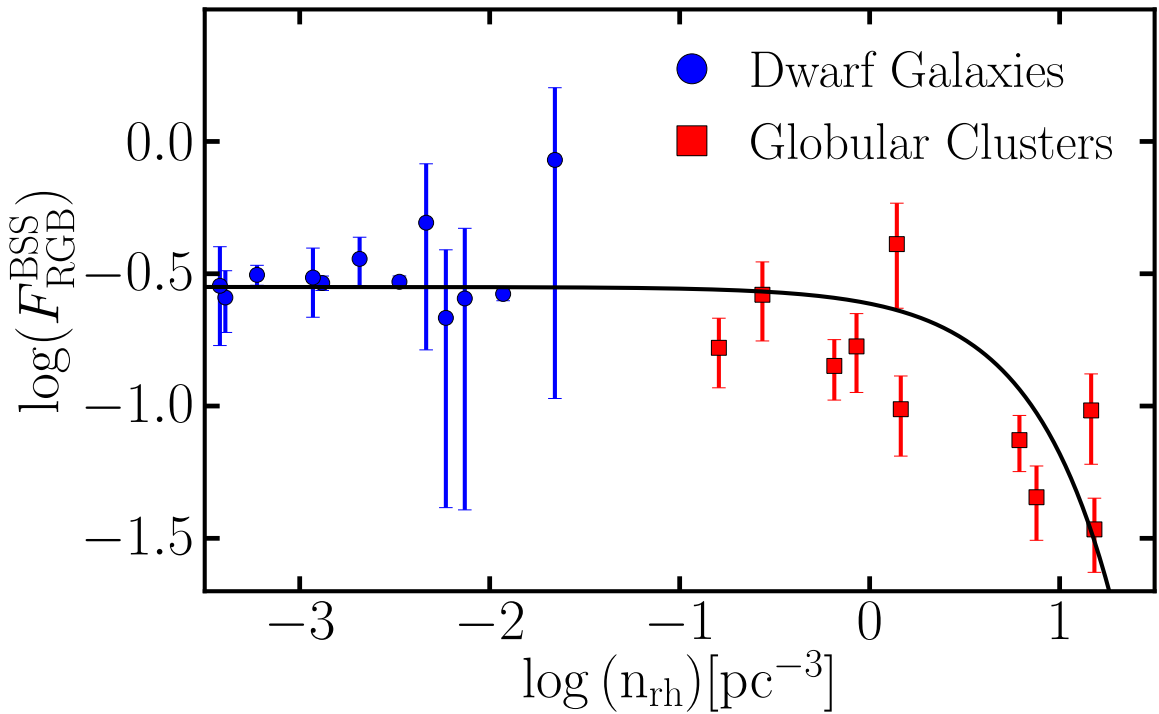}
\caption{Specific fraction of blue stragglers $F_{\rm{RGB}}^{\rm{BSS}}$ is plotted against 
density, calculated inside one half-light radius of each system.
 While dwarf galaxies show a flat distribution on the low
 density regime, clusters show an anti-correlation in the high density regime. 
The function fitted is shown as a solid black line, which corresponds to
 $\log({F_{\rm{RGB}}^{\rm{BSS}}})$ $\propto$ ($-0.063\pm0.007$) $n_{\rm{rh}}$.
}
\label{freq_density}
\end{figure}

\begin{figure}
\plotone{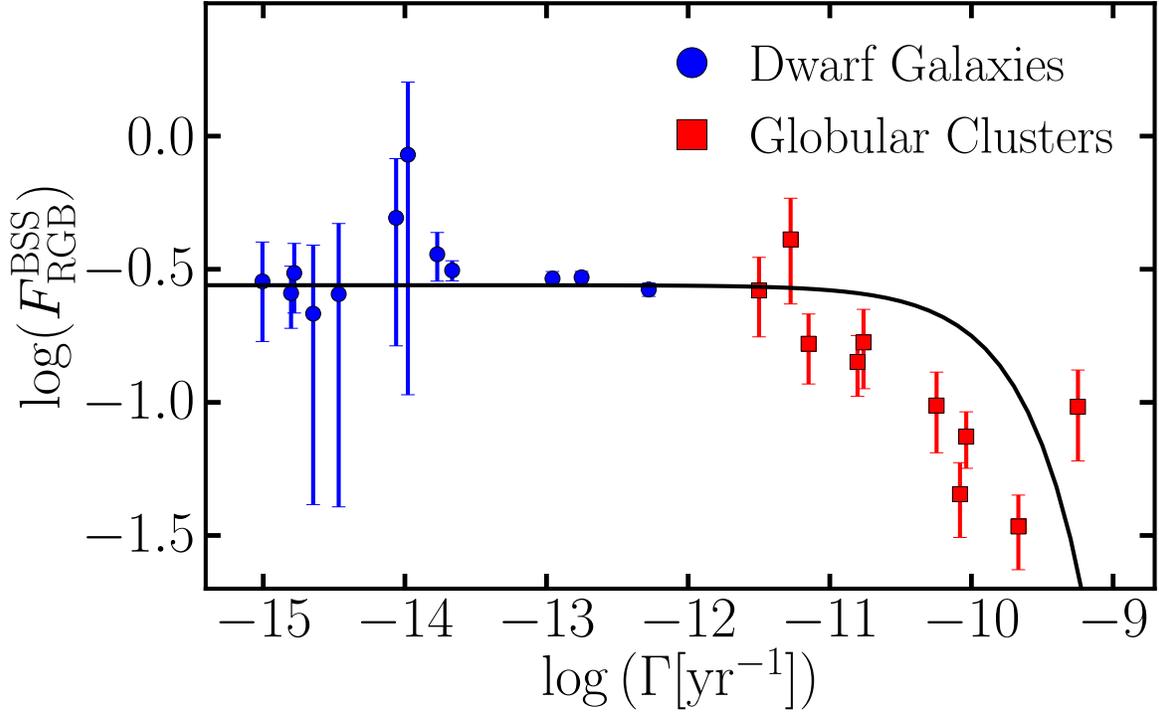}
\caption{Specific fraction of blue stragglers $F_{\rm{RGB}}^{\rm{BSS}}$
 against the rate for single-single star encounters,
 as calculated in \citet{leigh11b}. The fitted function is
$\log({F_{\rm{RGB}}^{\rm{BSS}}})$ $\propto$ ($-1.9\pm0.2$)$\times10^{9}$ $\mathit{\Gamma}$ 
and is illustrated as a solid black line.
}
\label{freq_gamma}
\end{figure}

Finally, we plot blue straggler specific frequencies against both the density within $r_{\rm{h}}$
 (see Figure~\ref{freq_density}) and the encounter rate between single-single stars 
(see Figure~\ref{freq_gamma}), as calculated in \citet{leigh11b}.
Figure~\ref{freq_density} shows that blue straggler frequencies of {\it all} our systems follow a 
single exponential trend with density, displaying a smooth transition between clusters and
galaxies.
Figure~\ref{freq_gamma}, on the other hand, shows that the same behavior is followed by 
the frequency of blue stragglers versus the encounter rate. The fraction of 
blue stragglers stays constant and high in the low density/low encounter 
rate regime spanned by our dwarf galaxies, and decreases with density 
and encounter rate in the range spanned by our globular clusters.
The fitting functions that describe the blue straggler specific frequency 
against these two parameters are:
\begin{equation}
\label{eq:freq_density}
\log{(F_{\rm{RGB}}^{\rm{BSS}})}=(-0.063\pm0.007)\,n{\textrm{[str/pc$^{3}$]}} + (-0.55\pm0.01) 
\end{equation}
\begin{equation}
\label{eq:freq_gamma}
\log{(F_{\rm{RGB}}^{\rm{BSS}})}=(-1.9\pm0.2)\times10^{9}\,\mathit{\Gamma} + (-0.56\pm0.01)
\end{equation}

\subsection{Blue straggler/Young Stars Discrimination}

By definition, blue straggler stars live in a region of the CMD that could also 
be inhabited by young stars. 
In old systems without recent episodes of star formation, like globular clusters have
traditionally been considered, the identification of blue straggler stars in the CMD is 
straightforward.
However, for satellites where recent episodes of star formation cannot be ruled out 
a priori, it is not immediately clear whether an observed extension of the MS
beyond the older turnoff is due to blue straggler or young stars.

We studied the numbers and magnitude distributions of stars inhabiting the region of
the CMD occupied by blue stragglers.
Based on these values, we estimate the ages and fractions of young stars that would
reproduce our observations in the absence of genuine blue stragglers. 
In this way, we can assess the likelihood that recent bursts of star formation could
be responsible for the stars observed beyond the main-sequence turnoff.

\begin{figure}
\plotone{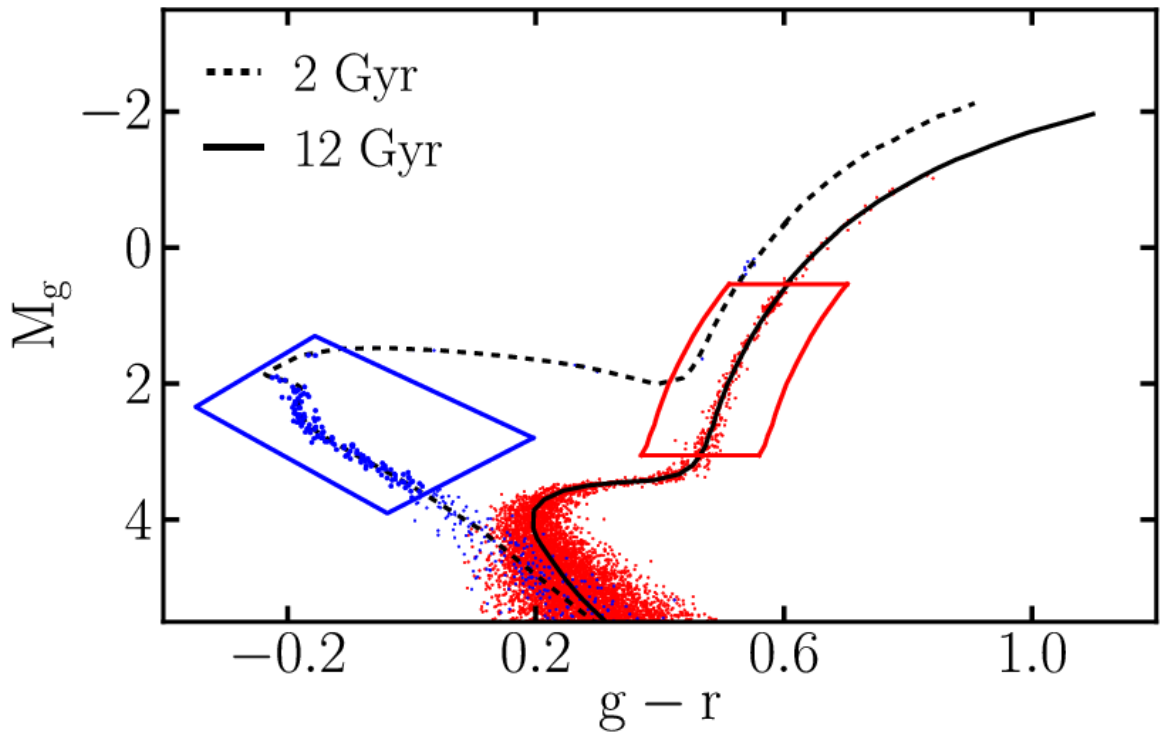}
\caption{Simulated CMD for fake stars with a fraction of young stars of $0.02$. 
Red and blue points represent, respectively, $2$\,Gyr and $12$\,Gyr stars. Solid and 
dashed black lines show the theoretical isochrones used, while blue and red boxes show 
the regions where blue straggler stars and RGB stars were counted. 
} 
\label{sim_CMD}
\end{figure}

Given the extremely low number of stars present in our dwarf galaxies, the only region of
the CMD that we can use to compare blue stragglers and young stars is the region 
previously defined as our blue straggler box, since all the other regions of the CMD
would show negligible number of blue stragglers and/or young stars compared to the
main old population or contamination stars (see Figure~\ref{sim_CMD} for an example
of the expected appearance of the CMD of an object where young stars could reproduce
the number and distribution of stars observed beyond the main-sequence turnoff in our
dwarf galaxies).

\begin{figure}
\plotone{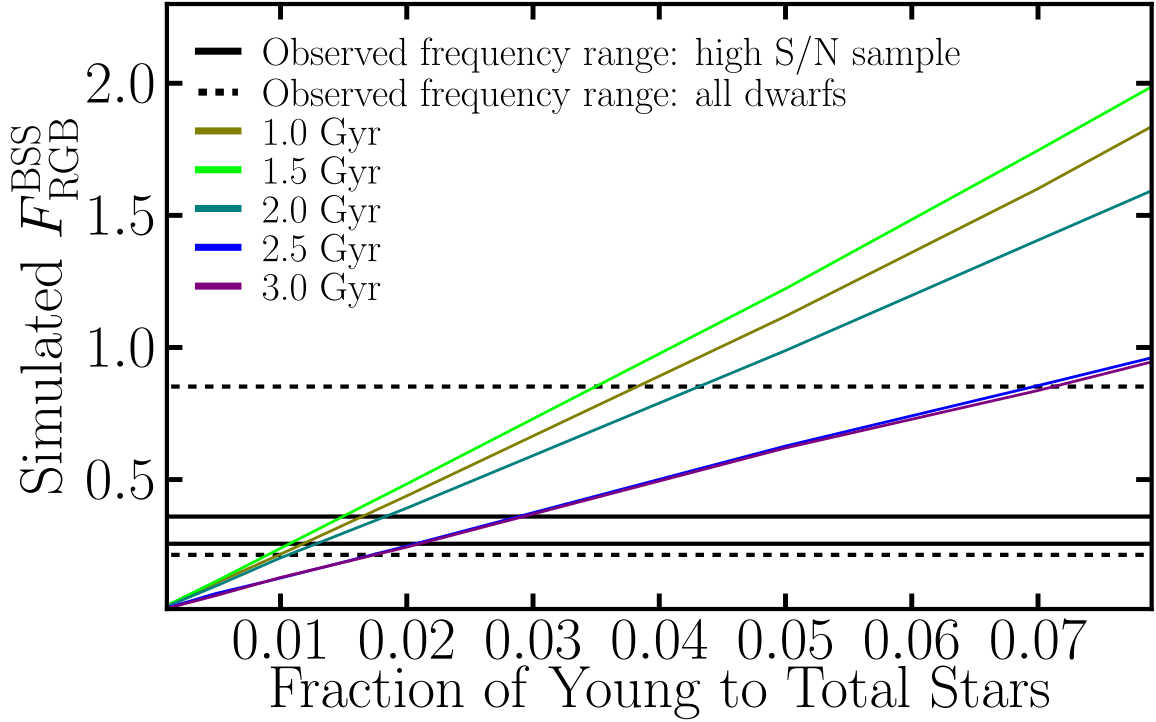}
\caption{ Simulated fraction of blue straggler stars corresponding to each young star
fraction, for different ages of the generated stars. 
The black dashed lines show the blue straggler fraction range observed in our 
complete galaxy sample.
The black solid lines show the blue straggler fraction range observed 
in the galaxies excluding the four systems with the highest frequency uncertainties.
} 
\label{diff_ages}
\end{figure}

To estimate the properties of the young stars that could reproduce the blue straggler
frequencies observed in our galaxies, we ran simulations where we generated both a
young and an old single stellar population. To generate them, we used two Padova
isochrones: a young one with an age varying from $1$ to $3$\,Gyr and an old one of 
$12$\,Gyr, both with an abundance of [Fe/H]$=-2.0$ which represents the average among 
the galaxy sample\footnote{Varying the metallicity of the isochrone introduces only minor
changes in our results, and therefore, for simplicity, we chose to keep the metallicity
constant.}. We also used the corresponding theoretical luminosity functions, based
on a \citet{chabrier03a} initial mass function, incorporating magnitude uncertainties 
consistent with our photometric data.
Once we populated the fake CMDs, we counted blue straggler-like and RGB stars in the 
same way we did for the real data.
Thus, for a given age $a$ and fraction of young stars $f$ we obtained a simulated blue 
straggler fraction $F(a,f)$ corresponding to the one that a given system with no genuine 
blue stragglers would show. 
By comparing the frequencies measured in the real data with the simulated data, we
obtained the fraction of young stars that would be needed to
mimic the observed population of blue stragglers.
In Figure~\ref{sim_CMD} a simulated CMD is shown for illustration,
with young and old populations of $2$ and $12$\,Gyr respectively, 
and a fraction of young over total stars of $0.02$.
The results of the simulations are shown in Figure~\ref{diff_ages}. This plot shows 
the fake blue straggler frequencies corresponding to different fractions of young stars, 
for ages ranging from $1$ to $3$\,Gyr. Also shown here are the ranges of blue straggler
 fractions actually observed: one including all the objects and the other excluding the
 four (out of twelve) galaxies with the largest frequency uncertainties.

Additionally, we constrained the age of young stars that could mimic 
blue stragglers by comparing the magnitude distribution of each set of young stars 
with those of the observed blue stragglers. 
We also used the globular clusters as a ``control sample".
Given that the number of stars in the UFDs are extremely low, to make the comparison 
statistically meaningful, we analyzed the magnitude distribution of the UFDs as a single 
group. 
The three classical dSphs in our sample were studied individually.
We carried out a Kolmogorov-Smirnov (KS) test to compare the different sets of stars,
and found that stars with ages of $2.5\pm0.5$\,Gyr were the only ones even marginally 
consistent with the magnitude distribution of the observed blue stragglers in our dwarf 
galaxies.
The magnitude distributions of blue stragglers in globular clusters are also consistent with
the one of $2.5$\,Gyr old stars and blue stragglers from dwarf galaxies.  
This result comes as no surprise.
For populations older than $\sim2.5$\,Gyr, turnoff stars leave 
what we defined as our blue straggler box progressively closer to its faint end while
younger populations will extend beyond the upper luminosity limit observed for blue
straggler, both in clusters and in galaxies. 

What is left to determine is the fraction of stars with ages in the range of 2--3\,Gyr that
would reproduce the specific fractions of blue stragglers observed in our dwarf galaxies.
Figure~ \ref{diff_ages} shows that to reproduce the lowest observed fraction of blue
straggler stars in all dwarf galaxies, a minimum young star fraction of $\sim1$--$2$\%
is needed.
For the upper limit, the fractions needed are $\sim4$--$7$\% for $2.5\pm0.5$\,Gyr
old stars.

In summary, the stars in the region of the CMD occupied by blue stragglers in our dwarf
galaxies can be attributed to recent bursts of star formation only if all the dwarf galaxies
in our sample formed stars $2.5\pm0.5$\,Gyr ago, and these stars account for
 $\sim1$--$7$\% of the total number of stars, or $\sim1$--$9$\% in mass fraction.
Furthermore, if we exclude the four systems with the highest blue straggler
frequency uncertainties, the fine-tuning of the star formation history of galaxies
 would have to be even greater to explain blue stragglers, since the young star fraction 
needed would have to be in the narrow range of $1$ to $3$ \%, which (as explained in the
discussion section) we deem highly unlikely.

\subsection{Radial Distribution Analysis}

We explored an additional line of evidence to help elucidate the nature of the blue straggler 
candidates in our Galactic satellites: we compared their radial distributions to those
of RGB and MS stars. How blue stragglers are distributed throughout a system can be
the result of a complex interplay between dynamical history and the dominant blue 
straggler formation mechanism. 
In our dwarf galaxy sample collisions are negligible and $2$-body relaxation times
are longer than the age of the universe and therefore dynamical evolution (mass
 segregation) is not expected. In this scenario, there is no reason to presume
central concentration of blue stragglers. Young stars, on the other hand
tend to be centrally concentrated with respect to other stellar 
populations in dwarf galaxies \citep[e.g.][]{harbeck01a,grebel01a}, and thus the radial 
distribution of the stars we classified as blue stragglers can help us distinguish 
genuine blue stragglers from young stars. 
In the case of the globular clusters in our sample, where we can assume a
priori that blue stragglers are genuine, eventual central concentration 
could shed some light into the relevance of collisions as a formation mechanism.

\begin{figure*}
\plotone{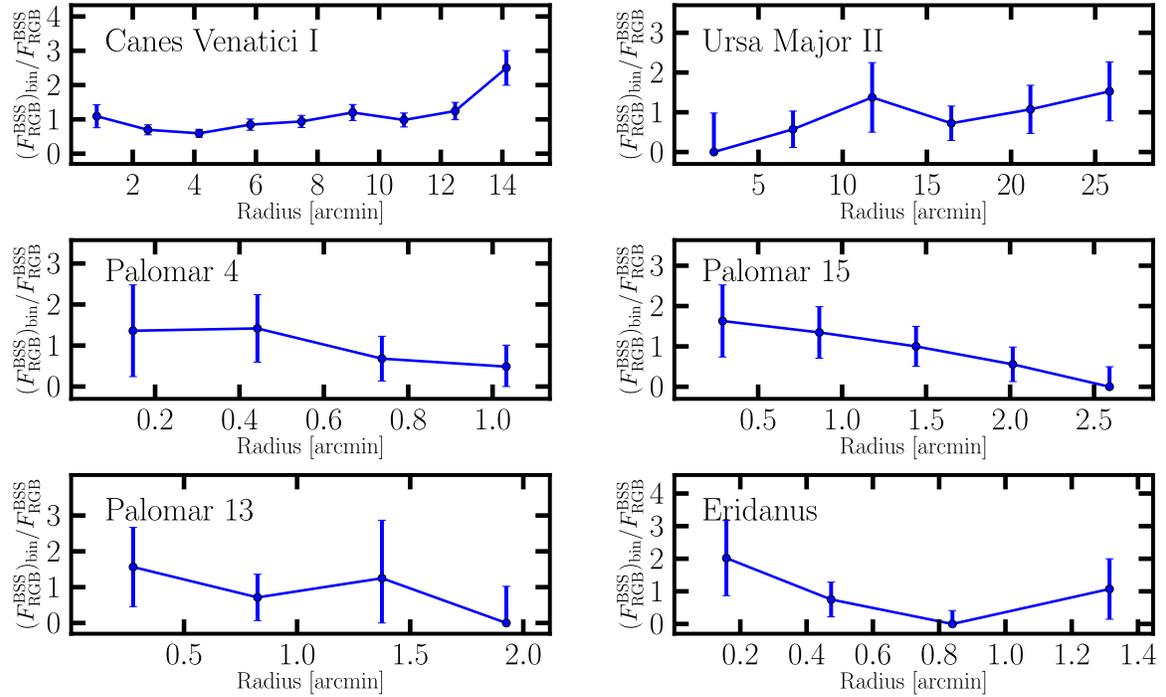}
\caption{Radial distribution of blue stragglers with respect to the one of RGB stars.
For each system, the curve shows the specific frequency of blue stragglers at different radii
normalized to the total specific frequency of blue stragglers.
All the objects where the radial distribution of blue stragglers were not statistically 
consistent with the one of RGB stars are plotted. 
{\it Top}: Two dwarf galaxies where blue stragglers are located preferentially 
on the outskirts of the system.
{\it Middle}: Two globular clusters where blue stragglers are centrally concentrated.
{\it Bottom}: Globular cluster Palomar~13 (left panel) with a central concentration of
blue stragglers and Eridanus globular cluster (right panel) where a bimodal distribution can 
be present.}
\label{diff_raddist}
\end{figure*}

For most satellites in our sample, we found that the radial distribution of blue stragglers 
is nearly indistinguishable from that of RGB stars.
Significant differences are seen only in $6/22=27\%$ of our systems. These objects
are: the UFDs Canes Venatici~I 
and Ursa Major~II and the globular clusters Palomar~4, Palomar~13, Palomar~15
and Eridanus.
Figure~\ref{diff_raddist} shows the blue straggler fraction versus radius, normalized
by the overall fraction of blue stragglers, for these six objects.
It is interesting that for Canes Venatici I and Ursa Major II, blue straggler stars are 
located preferentially in the outer regions. 
This behavior is also observed in galaxies like Draco, although for this galaxy 
the difference is too small for the KS test to differentiate both radial distributions.  
For the globular clusters in the figure, a clear radial concentration is observed 
(except for Eridanus, where a bimodal distribution might be present). 

It is worth reminding the reader that the area we used to select blue stragglers
 corresponds to twice the half-light radii of the systems, and therefore features in the
radial distribution present at larger distances will not be observed if the objects extend
much further than this. However, we do not anticipate this to be a problem given the
extremely low densities at radii larger than $2\times r_{\rm{h}}$. 

\clearpage{}

\section{Discussion}

\subsection{Dwarf Galaxies}

In both classical and UFD galaxies, blue stragglers are ubiquitous, regardless
of how low the stellar densities or encounter rates are. 
Their specific frequencies are
high compared to those observed in globular clusters and are found
to be statistically consistent with being constant over a six magnitude range.
Given that we found the RGB populations to scale linearly with luminous mass, 
this is equivalent to saying that the number of blue stragglers grows almost
linearly with the total stellar mass of the system.

We used simulations of young populations to compare their 
photometric properties with those of the blue stragglers observed 
in dwarf galaxies and conclude that the latter are genuine, as opposed to young stars. 
A number of facts support this conclusion:
(1) for young stars to have magnitude distributions statistically consistent 
with the blue stragglers observed in dwarf galaxies, 
their ages need to be closely clumped around $2.5$\,Gyr. 
This result can be readily understood when we consider that our brightest blue 
stragglers have an absolute magnitude of $M_{\rm{g}}\sim1.9$, coincident with the
 magnitude at which a $2.5$\,Gyr old star evolves out of our blue straggler star box.
This is also the magnitude corresponding to a star with twice the mass of a turnoff star
$12$ to $13$ Gyr old, an expected result if we are seeing blue stragglers formed by
collisions (either single-single or in binaries) or mass-transfer in binary systems.
(2) The magnitude distributions of blue stragglers
in both dwarf galaxies and globular clusters (where they can be reliably classified as blue
stragglers) are completely consistent.
(3) The lack of central concentration of blue stragglers in dwarf galaxies is 
consistent with the scenario wherein these stars form from mass-transfer or mergers
in primordial binaries or multiple systems, rather than being the result of a recent star
formation episode. In the latter case the young stars would be expected to be located
preferentially near the central regions.
(4) From the simulations we also determined that to reproduce the range of observed 
blue straggler frequencies, the $2.5\pm0.5$\,Gyr old stars should constitute 
$1$--$7$\% of the total number of stars in all the dwarf galaxies in our sample, an 
unlikely fine-tuned common star formation history.

Most galaxies in our sample have half-light densities of $10^{-2}$--$10^{-3}$\,stars 
pc$^{-3}$, i.e., at least $10$ times less dense than the solar neighborhood 
\citep{latyshev78a}. 
Given these extremely low stellar densities, blue stragglers formed by collisions between 
stars can be safely ruled out. When considering collisions between single, binaries
or triple stars, the collision times (calculated as in \citealt{leigh11b}) are orders
of magnitude higher than the age of the universe. Even though some physical processes
have been particularly successful in explaining collisions in low density environments,
they might not explain the blue stragglers observed in systems like our dwarf galaxies.
For instance, the triple evolution dynamical instability proposed by 
\citet{perets12a} produces encounter rates which are too low in systems with low
numbers of stars to explain our dwarf galaxy blue stragglers.

If collisions of any kind cannot account for blue stragglers in our systems, their presence
can only be explained if they formed via mass-transfer and/or mergers in primordial
 binaries, whether or not they have more companion stars that are members of the system. 
Two powerful correlations were found to support this claim. 
When plotting blue straggler fractions against both density and encounter rate,
we found that a single exponential function could reproduce the behavior of all 
satellites in our sample. In this context, dwarf galaxies live in the lower 
density/encounter rate regime, displaying high and similar values of blue straggler
fractions. This points to the fact that collisions neither significantly create nor prevent the
 formation of their blue stragglers.
On the other hand, as we explain below, close encounters in higher
density environments prevent blue straggler formation by altering the configuration 
of the binary or multiple systems.
Finally, the lack of central concentration of blue stragglers in all our dwarf galaxies
is also consistent with the binary/multiple system scenario, implying that these stars
can be formed in all regions of our galaxies and not just their slightly higher density central
regions.

The similarity in the blue straggler fractions observed in galaxies can be explained if the
 primordial binary star fractions are also similar. 
While this should be further confirmed by observations, hints that this is in fact the
case already exist (\citealt{geha13a} measured the binary fraction of two
ultra faint galaxies, finding identical binary fraction values of $47$\%).

\subsection{Globular Clusters}

Our observations show that, for the globular clusters in our survey, 
the specific frequency 
of blue stragglers decreases when there is an increase in a particular physical parameter
of the host system, such as luminosity, stellar densities,  encounter rate
and total stellar mass.
The anti-correlation between $F^{\rm{BSS}}_{\rm{RGB}}$ and the luminosity of the systems
is similar to the one observed for inner halo clusters, even though our
globular clusters are on average less luminous and larger ($5$--$10\times$). 
The slope of our anti-correlation is consistent within the uncertainties with the 
one found by \citet{piotto04a} using a sample of $56$ globular clusters with 
$-6>M_{\rm{V}}>-10$, and by \citet{sandquist05a}, which extends the results of
 Piotto et al. results with lower luminosity clusters down to $M_{\rm{V}}\sim-4$.
The anti-correlation derived in our study extends the existing ones to absolute
magnitudes as faint as $M_{\rm{V}}\sim-2.5$. At this faint end, the fraction 
of blue straggler stars in our globular clusters is comparable to that of dwarf galaxies.
Despite the consistency between our correlations, 
there is a key difference between our results and the one by \citet{piotto04a}: 
we study blue stragglers within $2 \times r_{\rm{h}}$ of our clusters, which
represents a significant fraction of the total cluster area, whereas Piotto's work focused 
on the cluster's cores.
This difference is important since, as proposed by \citet{leigh11c}, the systems with the
higher relaxation times/higher mass would not have had time to sunk their blue stragglers
to the innermost regions by two-body relaxation. 
This would reduce the number of blue stragglers, $N_{\rm{BSS}}$, found in the high mass
clusters when counting them in the most central regions, but that would not affect
the trend of $N_{\rm{BSS}}$ when the region of the cluster considered represents a 
considerable fraction of the total cluster area.
Thus, if dynamics in the central regions do not destroy the progenitors of blue stragglers
and these stars are homogeneously formed within the clusters, mass segregation would
translate to a sublinear dependence of $N_{\rm{BSS}}$ with cluster mass enclosed when
studying only the central regions whereas a linear dependence  would be expected when 
considering larger areas. Our Figure~\ref{number_mass} then argues against mass
segregation playing an important role in the blue straggler
counts in our globulars.

As was the case for dwarf galaxies, collisions alone are unable to explain the fraction of 
blue stragglers observed in our globular clusters.
Based on the collision times, calculated as in \citet{leigh11b}, collisions between single,
binary or triple systems can account only for a small contribution to the blue straggler
numbers observed in our highest density clusters. 
Thus, there should be another dominant blue straggler
 formation mechanism at work.
Figures~\ref{freq_density} and ~\ref{freq_gamma} show that globular clusters inhabit
our higher density, higher encounter rate regime, showing a systematic decrease
with both physical parameters. We interpret these trends as supporting a scenario 
where mass-transfer or mergers in binary or multiple star systems are the dominant 
blue straggler formation mechanism in the outer halo globular clusters.
The following pieces of evidence support this scenario:
(1) The behavior of the frequency of blue stragglers
is well fitted by single trends with smooth transitions between dwarf galaxies and 
clusters, which points to a common origin for their blue stragglers.
(2) Systematic decrease of blue straggler fraction with encounter rate and density is
inconsistent with the collision scenario. Instead, this points to encounters preventing 
blue straggler formation in our globular clusters.
(3) The expressions shown in Equations~\ref{eq:freq_density} and \ref{eq:freq_gamma}
describing the exponential decay of the frequency of blue stragglers with both density
and encounter rate arises naturally if the relative decrease in the fraction of blue stragglers
goes as the ratio between the age of the system and the collision time.
It is worth pointing out that collisional blue stragglers might have shorter lifetimes
than systems formed through mass-transfer \citep{chatterjee13a}.  
This means that we cannot rule out the possibility that a fraction of blue stragglers
in denser systems still formed through collisions involving binaries 
(that would have otherwise undergone mass-transfer to form a blue straggler)
and that they quickly evolved away from the blue straggler region.
We argue that this would be only a second order effect, 
because the differences in the lifetimes of blue stragglers produced
by the different mechanisms is much less than the one needed to explain the decline
in the frequency of blue stragglers observed for our clusters.

Aside from our study, there is mounting evidence favoring a binary origin for blue
stragglers. 
A direct link between blue straggler stars and binaries has been determined by
 \citet{preston00a}, who derived a binary fraction of $68$\% among their metal-poor
field blue straggler stars, and \citet{mathieu09a} who estimated a binary fraction
of $76$\% among blue straggler stars in NGC 188.
Palomar~13, one of the clusters with the highest blue straggler frequencies is known
to have a relatively high fraction of binary stars, $30\pm4$\% \citep{clark04a}, 
and many of their blue straggler stars were proved to show 
significant velocity variations, suggesting these are unresolved binary systems 
\citep{bradford11a}.

In our globular cluster sample,
the radial distributions of blue stragglers are in most cases 
indistinguishable from those of RGB or MS stars, consistent in principle 
with the binary scenario. 
However, a clear central concentration of blue stragglers is observed in a few clusters: 
Palomar~4, Palomar~13  and Palomar~15, while a bimodal distribution 
might be present in Eridanus. 
At first glance, this may seem contradictory with our
interpretation of Figure~\ref{freq_density} that higher density environments favor
the destruction or separation of binary or multiple systems progenitors
of blue stragglers, but the trend of frequency with density followed by different
objects should not necessarily be expected to hold within a {\it single} system.
\citet{fregeau09a} studied the evolution of binaries in dense stellar systems
and found an increase with time of the core binary fraction, which could be understood as 
a consequence of a complex interaction between mass segregation of binaries into the 
core, and their subsequent destruction there.
In addition, once formed, blue stragglers can also migrate toward the central
regions through mass segregation. 
In summary, dynamical processes likely to occur in globular clusters
severely complicate the interpretation of the trends observed within an 
individual object.

\section{Conclusions}

We have presented a comprehensive analysis of the blue straggler star population in a 
representative sub-sample of Galactic outer halo satellites. 
This photometrically homogeneous sample includes ten low density globular clusters,
 three classical dSph galaxies and nine of the recently discovered UFD galaxies. 
Despite their diverse physical properties, all these satellites are relatively loose and scarcely
populated when compared to inner halo globular clusters, where most blue straggler star
studies have been carried out. 
Given the extremely long collision times of our systems, 
collisions involving single, binary or triple stars can only account for a small fraction of 
the blue stragglers of our highest density clusters, while their influence on dwarf galaxies 
should be negligible.
Our sample provided an opportunity to study blue straggler populations in a new
 density/luminosity regime. We claim that the dominant blue straggler formation 
mechanism in these type of systems is mass-transfer or mergers in binary or multiple
star systems.
In the higher encounter rate regime spanned by our globular clusters, encounters 
prevent blue straggler formation by altering the configuration of the star systems that
would otherwise produce blue straggler stars.

Our results can be summarized as follows:

1. We found blue stragglers to be ubiquitous among globular clusters and dwarf galaxies,
including the UFDs.

2. The blue straggler populations in both classical dSphs and UFDs show a remarkably 
high and constant distribution of their fractions over an absolute magnitude range of 
more than six magnitudes, and a density range of two orders of magnitude.

3. The behavior of the frequency of blue stragglers is well fitted by single trends with 
smooth transitions between dwarf galaxies and clusters, which points to a common 
origin for their blue stragglers. 

4. The fraction of blue straggler stars is high and flat in the extremely low encounter rate
 regime spanned by dwarf galaxies, while it decreases exponentially with increasing stellar
 density or encounter rate for the regime spanned by our outer halo globular clusters.

5. There is a well-defined anti-correlation between the fraction of blue straggler stars
and absolute magnitude for the outer halo clusters in our sample. This trend has 
already been observed in inner halo clusters and it is also interpreted as a consequence of 
the binary origin of the blue straggler population.

6. Comparing the magnitude distribution of the observed blue stragglers in dwarf galaxies
with those of simulated single stellar populations, we find that for blue stragglers 
in dwarf galaxies to be young stars, they would have to correspond
to a $2.5\pm0.5$\,Gyr old population. In addition, to match the observed blue straggler 
fractions seen in galaxies, young stars would have to comprise between $\sim1$--$7$\%
 of the total number of stars.
Such fine-tuned requirements make it unlikely that we are mistakenly classifying young 
stars as blue stragglers. 

7. The radial distribution of blue stragglers in most objects is statistically 
consistent with the ones found for their RGB and MS stars. Only a few exceptions are 
found, notably the central concentration seen in Palomar~4, Palomar~13, Palomar~14 
and the bimodal distribution in Eridanus. In all these cases, dynamical processes, like
mass segregation, are likely to alter the primordial binary population and therefore the interpretation of the trends observed within individual objects is not straightforward.

We thank the referee for his very useful comments that helped improve
this paper significantly.
F.~A.~S.~ thanks Andr\'es Guzm\'an for useful discussions.
F.~A.~S.~acknowledges support from CONICYT-PCHA/Doctorado 
Nacional/$2010$-$21100133$.
R.~R.~M.~acknowledges partial support from CONICYT Anillo project ACT-1122 and 
project BASAL PFB-$06$ as well as from the FONDECYT project N$^{\circ}1120013$. 
M.~G.~acknowledges support from the National Science Foundation under award number 
AST-0908752 and the Alfred P.~Sloan Foundation.
S.~G.~D.~ was supported in part by the NSF grant AST-$0909182$ and by the Ajax Foundation.
This work was supported in part by the facilities and staff of the Yale University Faculty 
of Arts and Sciences High Performance Computing Center.

\end{document}